\documentclass[conference]{IEEEtran}
\IEEEoverridecommandlockouts

\usepackage{cite}
\usepackage{amsmath,amssymb,amsfonts}
\usepackage{algorithmic}
\usepackage{graphicx}
\usepackage{textcomp}
\usepackage{xcolor}

\usepackage{mathptmx}
\usepackage{booktabs}
\usepackage{orcidlink}

\usepackage{hyperref}

\usepackage{setup}
\usepackage{url}
\usepackage{enumitem}

\usepackage{balance}

\def\BibTeX{{\rm B\kern-.05em{\sc i\kern-.025em b}\kern-.08em
    T\kern-.1667em\lower.7ex\hbox{E}\kern-.125emX}}

\usepackage{tikz}

\newcommand*\insidebox[1]{\tikz[baseline=(char.base)]{
            \node[shape=rectangle,draw,inner sep=2pt] (char) {#1};}}

\makeatletter 
\newcommand{\linebreakand}{%
  \end{@IEEEauthorhalign}
  \hfill\mbox{}\par
  \mbox{}\hfill\begin{@IEEEauthorhalign}
}
\makeatother 

\begin{document}


\title{Improving Merge Pipeline Throughput in Continuous Integration via Pull Request
Prioritization}

\author{
\IEEEauthorblockN{Maximilian Jungwirth}
\IEEEauthorblockA{\textit{BMW Group, University of Passau} \\
Munich, Germany \\
maximilian.jungwirth@bmw.de}
\and
\IEEEauthorblockN{Martin Gruber}
\IEEEauthorblockA{\textit{BMW Group}\\
Munich, Germany \\
martin.gr.gruber@bmw.de}
\and
\IEEEauthorblockN{Gordon Fraser}
\IEEEauthorblockA{\textit{University of Passau} \\
Passau, Germany \\
gordon.fraser@uni-passau.de}
}

\maketitle

\begin{abstract}
Integrating changes into large monolithic software repositories is a
critical step in modern software development that substantially impacts the speed of feature
delivery, the stability of the codebase, and the overall
productivity of development teams.
To ensure the stability of the main branch%
, many organizations use merge pipelines
that test software versions before the changes are permanently integrated.
However, 
the load on merge pipelines is often so high that they become
bottlenecks, despite the use of 
parallelization. Existing optimizations frequently rely on
specific build systems, limiting their generalizability and
applicability. 
In this paper we propose to optimize the order of PRs in merge
pipelines using practical build predictions utilizing only historical
build data, PR metadata, and contextual information to estimate the
likelihood of successful builds in the merge pipeline. By dynamically
prioritizing likely passing PRs during peak hours, this approach
maximizes throughput when it matters most.
Experiments conducted on a real-world, large-scale project
demonstrate that predictive ordering significantly outperforms traditional
first-in-first-out (FIFO), as well as non-learning-based ordering
strategies.
Unlike alternative optimizations, this approach is agnostic to the underlying
build system and thus easily integrable into existing automated merge pipelines.


\end{abstract}


\section{Introduction}

Monorepos have gained substantial popularity among software engineering teams
as they facilitate easy code sharing and thus reduce code
duplication; they provide centralized third-party dependency management,
mitigating the complexities of dependency hell; and they create a linear version
history, which supports reproducible integration testing and simplifies the
identification of changes that introduced
faults~\cite{levenberg2016why,harry2017largest,goode2014scaling,lucido2017uber,mens2024depencyhell}.
%
However, the centralized nature of monorepos also introduces significant risks.
A critical requirement for successful monorepo usage is maintaining a ``green''
mainline branch, where all builds pass
successfully~\cite{ananthanarayanan2019keeping}.
When developers branch off from a broken mainline, they may find themselves
fixing the same issues repeatedly—issues that they did not introduce—leading to
wasted effort and reduced overall efficiency.

Continuous integration (CI) practices address this challenge.
However, traditional pre-merge testing that runs on feature branches does not
guarantee that the same tests are still passing after the merge, as it neglects
conflicts between concurrent changes happening on the mainline branch after the
initial branch-off point. This limitation has led to the development of merge
pipelines, which test incoming changes as they will be merged. Merge pipelines
have been adopted by large organizations (Uber~\cite{ananthanarayanan2019keeping}, AirBnB~\cite{kudelka2022evergreen}, Aviator~\cite{jain2023aviator}) and are supported by various CI systems (GitLab~\cite{gl-merge}, GitHub~\cite{gh-merge},~Zuul~\cite{Zuul}).

\begin{figure}
  \centering
  \includegraphics[width=.45\textwidth]{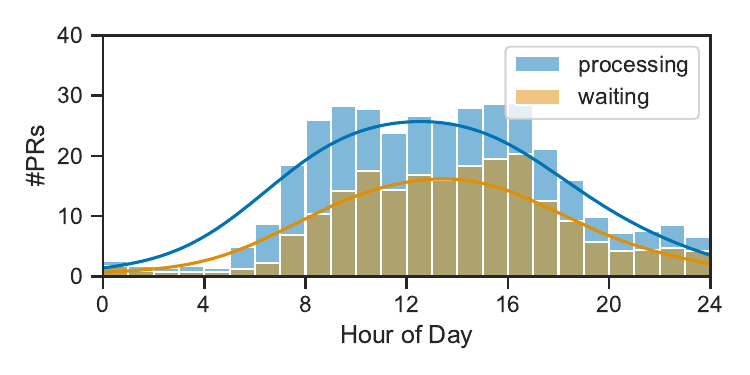}
  \caption{
      Average number of processing and waiting PRs in the merge pipelines of our case study throughout a day.}
  \label{fig:pr_over_day}
\end{figure}

While merge pipelines help maintaining a green mainline, they introduce new
challenges. In large organizations, hundreds or even thousands of changes need
to be merged every day, all passing through the merge pipeline, making it a
potential bottleneck~\cite{ananthanarayanan2019keeping}. Resulting delays are
especially costly during business hours when developer activity is at its
peak~(see~\cref{fig:pr_over_day}).

Existing merge pipeline optimizations are often limited by the success rate of
builds or require deep integration with the build
system~\cite{ananthanarayanan2019keeping}, which is why we propose a novel,
build tool-agnostic approach that improves existing merge pipelines by delaying
likely failing pull requests (PRs) when it matters most (\ie during periods of
peak developer activity) to increase throughput. We operationalize this approach
using a predictive model that leverages historical build data, PR metadata, and
contextual information to estimate the likelihood of a successful build in the
merge pipeline for each PR. By prioritizing PRs based on their predicted success
probability, our approach minimizes failed builds and enhances throughput,
particularly under high developer activity, when delays and bottlenecks are most
impactful. The approach itself is not tied to peak activity times, however its
benefits are most prominent during these periods. Likely failing PRs need to be
processed differently; as a simple initial mitigation we delay them to
non-business hours, ensuring they are handled overnight when feedback times are
not crucial, however, other strategies also are conceivable.

Our contributions include:
\begin{itemize}
  \item A detailed analysis of existing merge pipeline systems, which have
  emerged in practice but have received limited attention in research, and their potential limitations.
  \item The design and implementation of a probabilistic model for
  predicting build outcomes in a merge pipeline context to optimize throughput
  during business hours.
  \item A snapshot- and simulation-based evaluation of our predictive ordering
  method against traditional first-in-first-out (FIFO) approach and non-learning
  based ordering.
\end{itemize}
The results of the evaluation show significant improvements of the
pipeline throughput, which contributes to the overall goal of increasing
developer productivity.


\section{Merge Pipelines}


Among the different stages in the CI process, merging is especially
critical. Once a PR is approved and ready to be integrated, it is
vital to ensure not only compatibility against the main branch but
also against other PRs that may be merged before the current one. In
high-velocity development environments, multiple PRs are often
processed and merged nearly at the same time.  High-velocity
development environments are therefore facing the challenge that the
integration of one PR can potentially affect the outcome of
others. Ensuring that all such changes are compatible is essential to
maintaining stability in the main branch. This requires testing
new changes in the exact context in which they would be merged, i.e.,
a merge pipeline~\cite{ananthanarayanan2019keeping, gl-merge,
  gh-merge, Zuul}. Such testing guarantees that pre-merge checks are
not broken by concurrently tested PRs, i.e., the mainline of the
codebase remains stable at all time.

The simplest way to implement a merge pipeline is to sequentially test every
incoming PR against the current head, wait until it finishes, merge it or reject
it, and then proceed with the next one. However, this method proves ineffective
in fast-paced environments with long pipeline execution times. For example,
the monorepo containing BMW's driving dynamics and autonomous driving software
stack, which has an average testing time of 60 minutes per PR, meaning that a sequential merge pipeline would only allow   merging 24 PRs per day. This is far too low compared to the daily churn rate of approximately 400 PRs awaiting merging.
To handle this high volume of changes efficiently, some form
of optimization is necessary to improve the throughput and scalability of the
merge pipeline.

\subsection{Parallelization in Merge Pipelines}
To address the limitations of sequential testing in fast-paced development
environments, practitioners have developed horizontal and vertical
parallelizations for merge pipelines.


\subsubsection{Vertical Parallelization}
Vertical parallelization refers to concurrent testing of multiple PRs by stacking potential integration states on top of each other. The stacking can be further split into optimistic and pessimistic vertical parallelization.



\paragraph{Optimistic Vertical Parallelization}
\begin{figure}
  \centering
  \includegraphics[width=0.45\textwidth]{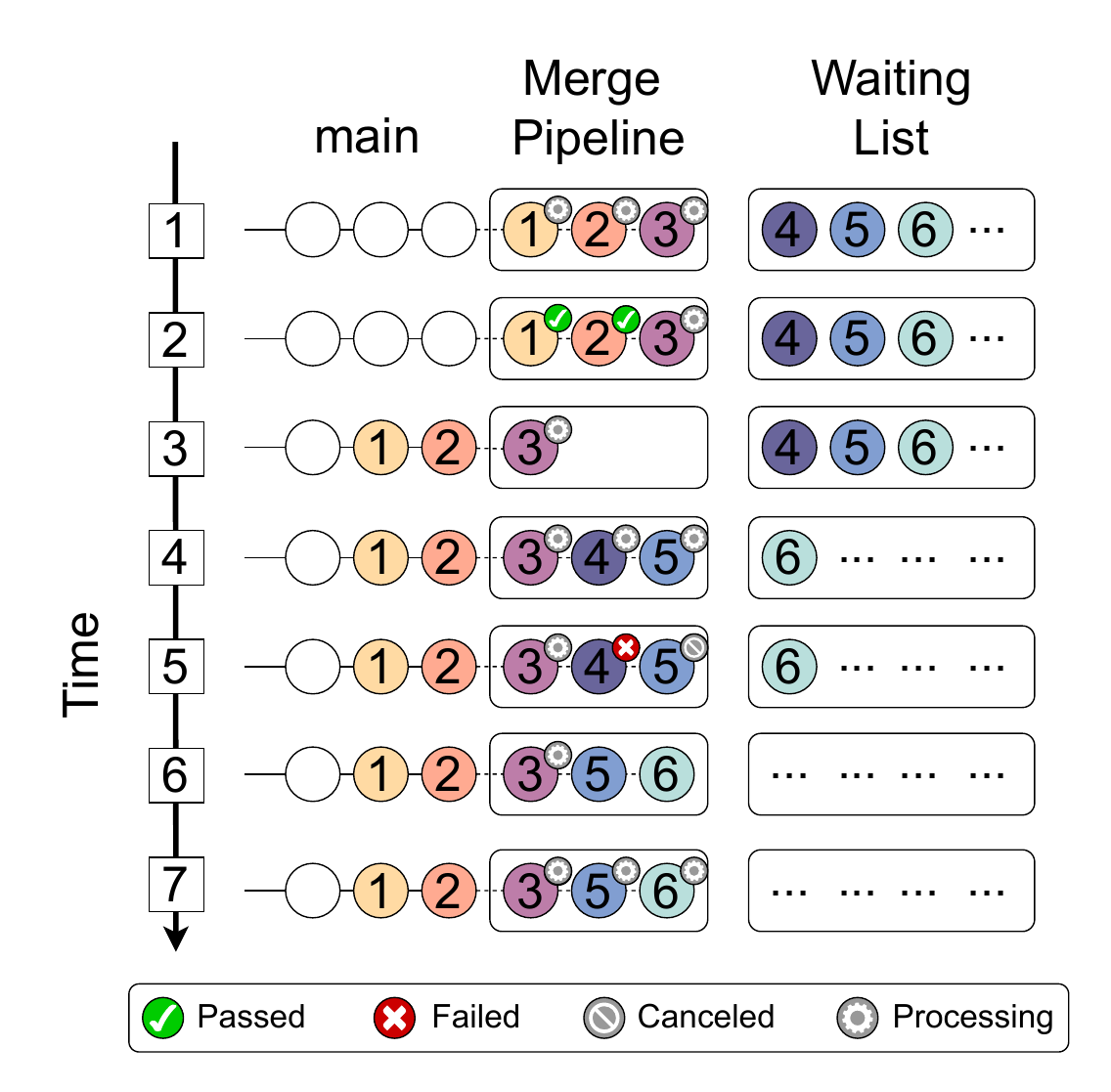}
  \caption{Illustration of optimistic vertical parallelization in a merge
  pipeline. Each PR speculates on the success of preceding PRs, leading to
  cascading failures when a PR fails.}
  \label{fig:optimistic}
\end{figure}

The CI system assumes that all pending changes in the merge pipeline will
succeed~\cite{gl-merge,gh-merge,Zuul}. \Cref{fig:optimistic} visualizes
optimistic vertical parallelization. In this setup, the merge pipeline assumes
that changes will integrate successfully until proven otherwise.
At~\insidebox{1} (time step one in~\cref{fig:optimistic}), the system processes
\textit{PR1}, \textit{PR2}, and \textit{PR3} in the pipeline. All three PRs run
their respective test steps concurrently
while speculating on the success of previous PRs (marked by dashed lines
in~\cref{fig:optimistic}). The system creates speculative integration states for
\textit{PR1}, \textit{PR2}, and \textit{PR3}. \textit{PR1} merges onto the
current head of the main branch and executes its build and test steps.
\textit{PR2} assumes that \textit{PR1} will integrate successfully and tests
itself on a speculative integration state, which includes \textit{PR1} merged
onto the main branch. Similarly, \textit{PR3} assumes the success of both
\textit{PR1} and \textit{PR2} and tests itself on a speculative integration
state that merges both preceding PRs onto the main branch. At \insidebox{2},
\textit{PR1} and \textit{PR2} complete their test steps successfully and clear
the way for merging into the main branch at \insidebox{3}. \textit{PR3}
continues its test steps without interruption, as already  speculated on the
success of \textit{PR1} and \textit{PR2}. At \insidebox{4}, the system enqueues
\textit{PR4} and \textit{PR5}, which start their processing in the available
slots. These PRs assume the success of all preceding PRs as well, including
\textit{PR3}. At \insidebox{5}, \textit{PR4} fails its test steps, which
triggers a cascading effect (which we call a \textit{merge pipeline reset}). The
pipeline cancels all the speculative integration states which speculated on the
success of \textit{PR4}, namely \textit{PR5}. At \insidebox{6}, the system
dequeues \textit{PR4} from the pipeline due to its failure. The system
reschedules \textit{PR5} to restart its processing from scratch and speculates
only on the success of \textit{PR3}. Additionally, the system enqueues
\textit{PR6} into the newly available processing slot to continue the pipeline's
progress. \textit{PR3} remains unaffected by the failure of \textit{PR4} and
continues its test steps without interruption, as it did not speculate
on the success of \textit{PR4}. At \insidebox{7}, the pipeline resumes normal
operations as \textit{PR5} and \textit{PR6} actively process. Real-world merge
pipelines typically have more processing slots than the three shown in the
example, emphasizing how failures can have severe cascading effects, especially
in environments with high failure rates or long-running pipeline
executions~\cite{ananthanarayanan2019keeping}.

\paragraph{Pessimistic Vertical Parallelization}
\begin{figure}
  \centering
  \includegraphics[width=0.35\textwidth]{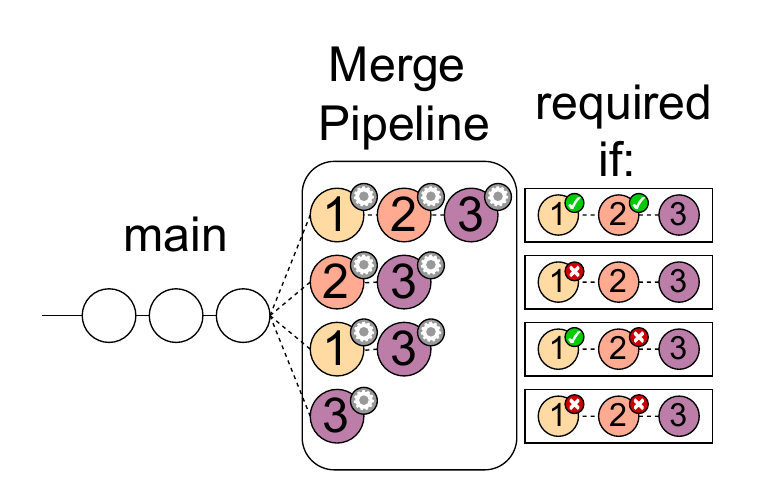}
  \caption{Illustration of fallback integration states in a merge pipeline. By
  precomputing integration states that account for potential PR failures, the
  system mitigates cascading failures and reduces latency, but at the cost of
  increased resource consumption.}
  \label{fig:pessimistic}
\end{figure}

Depending on the outcomes of the three PRs in the above example,
\cref{fig:pessimistic} illustrates all possible integration states that the
merge pipeline might execute for \textit{PR1}, \textit{PR2}, and \textit{PR3}.
To mitigate cascading effects of failures in merge pipelines,
Uber~\cite{ananthanarayanan2019keeping} proposed to maintain
fallback integration states and to process them concurrently.
This involves speculating not only on the successful integration of all PRs in the pipeline, but also accounting for scenarios where some PRs fail (as shown in \cref{fig:pessimistic}).
Each fallback path represents a distinct speculative integration state. For example,
in \cref{fig:pessimistic}, the system precomputes and maintains all fallback
states concurrently. If \textit{PR1} fails, the pipeline immediately transitions
to an integration state excluding \textit{PR1} (\ie the states that include
\textit{PR2} and \textit{PR3}, or, if \textit{PR2} also fails, the state with
only \textit{PR3}). By maintaining fallbacks, the system reduces latency and
improves throughput by avoiding to restart the merge pipeline from
scratch in the event of a failure.

However, this strategy introduces the risk of
resource inefficiency, as fallback states that are precomputed but ultimately
unused consume unnecessary computational resources. For instance, in
\cref{fig:pessimistic}, if \textit{PR1}, \textit{PR2}, and \textit{PR3} all pass
in the merge pipeline, the system only requires the three integration states
speculatively stacked onto each other. Any additional precomputed integration
states waste valuable resources. This inefficiency becomes apparent when
comparing the three concurrently running builds in \cref{fig:optimistic} at time
step two to the eight concurrently running builds in \cref{fig:pessimistic}. The
system must carefully balance the number of fallback paths with the availability
of computational resources and the typical failure patterns observed in the
repository~\cite{ananthanarayanan2019keeping}.

\subsubsection{Horizontal Parallelization}
While vertical parallelization processes multiple PRs by speculating on their
pipeline results, horizontal parallelization aims to increase throughput by
processing independent changes in parallel. This parallelization can be achieved
in a static or dynamic manner.
\paragraph{Static Horizontal Parallelization}
The simplest way of implementing horizontal parallelization is to statically
define multiple merge pipelines along the repository structure (\eg modules or
projects). By doing so, changes within different parts of the code can progress
independently, reducing contention and improving throughput for unrelated
changes. Statically split pipelines maintain a consistent structure, which
simplifies management and monitoring. However, this method may not optimally
handle cross-cutting changes (\eg cross-project breakages might slip through) or
fluctuations in project activity levels (\eg heavy load on an individual merge
pipeline).


\paragraph{Dynamic Horizontal Parallelization}
To overcome the limitations of static horizontal parallelization, an alternative
is to dynamically test PRs independently based on their affected
components or build targets. This strategy requires a build system that
organizes the software into targets, representing collections of source
files and their dependencies that collectively produce one or more executables
(\ie a test)~\cite{ananthanarayanan2019keeping}. Using this structure, the
system constructs an acyclic dependency graph to execute the build targets of a
software system in topological order, ensuring that build actions remain
hermetic and consistent~\cite{ananthanarayanan2019keeping}. Dynamic testing of
PRs begins by analyzing the affected build targets of incoming PRs, such as
modified source files or dependencies. The system then evaluates PRs
independently based on whether their affected targets overlap. If two PRs do not
overlap in their affected areas, the system tests them without speculating on
the outcome of other PRs.

\begin{figure}
  \centering
  \includegraphics[width=0.45\textwidth]{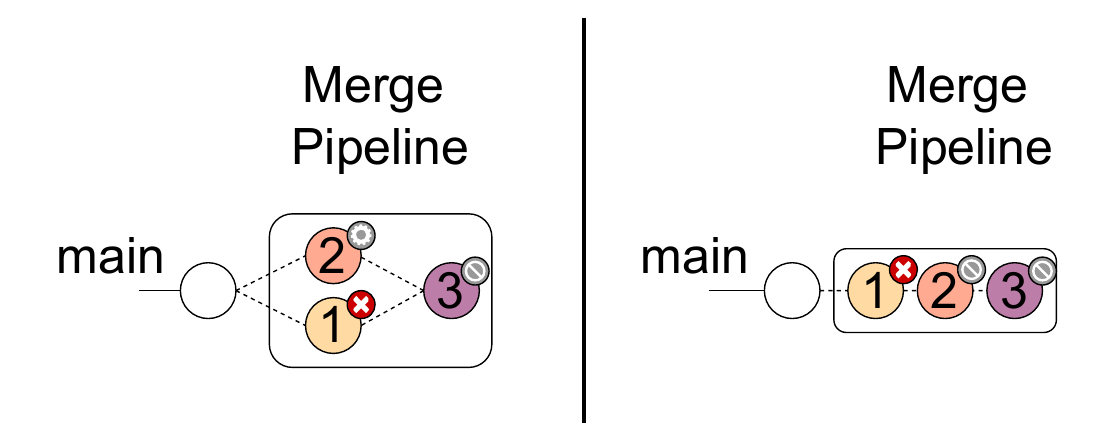}
  \caption{
    Comparison of PR testing strategies:
    (Left) Dynamic approach independently tests non-overlapping PRs using dependency analysis, minimizing rescheduling.
    (Right) Without dependency analysis, PR failures lead to unnecessary rescheduling of dependent and independent PRs.
  }
  \label{fig:dynamic}
\end{figure}

This approach enables fine-grained parallelization and
dynamically adapts to changing development patterns, thereby improving merge
pipeline efficiency. For example, consider \textit{PR1}, \textit{PR2}, and
\textit{PR3}. If the dependency analysis identifies that \textit{PR1} and
\textit{PR2} are independent, the system tests them in parallel without
speculating on the outcome of the preceding PRs. As shown in \cref{fig:dynamic}
on the left, the system tests \textit{PR1} and \textit{PR2} independently. If
\textit{PR1} fails, the system reschedules only \textit{PR3}, which depends on
\textit{PR1}. In contrast, when not performing independent testing
(depicted on the right of \cref{fig:dynamic}), the failure of \textit{PR1} would
require rescheduling both \textit{PR2} and \textit{PR3}, as their integration
states would depend on the outcome of \textit{PR1}. However, implementing this
strategy requires a build infrastructure capable of performing detailed
dependency analysis, which may not be available in all environments.
Additionally, the efficiency of this approach depends on the quality of the
dependency definitions (which must remain minimal) and the granularity of the
individual build targets.

\subsection{Existing Merge Pipeline Systems}
Several CI systems have implemented merge pipelines and their optimizations to
address the challenges of high-velocity development environments. GitHub's merge
queue and GitLab's merge train features support vertical parallelization with a
configurable static window size \cite{gh-merge, gl-merge}. This allows for
speculative merging and testing of multiple PRs simultaneously, improving
throughput while maintaining build stability.

Zuul~\cite{Zuul} uses a dynamic algorithm that adapts the
window size to the current success rate of changes. This
approach---inspired by the Transmission Control Protocol's flow
control mechanism---aims to minimize resource waste and increase
throughput~\cite{Zuul}. It initializes the window to a
default value, which is increased by a fixed value for each
successfully merged change, up to a predefined ceiling. When a PR
fails its testing steps, the system halves the window size down to a
predefined floor. This dynamic adjustment allows the window to
contract rapidly when changes begin to fail and expand gradually as
they succeed. Zuul~\cite{Zuul} additionally enables distributing load
horizontally into project-specific merge pipelines.

While these generic solutions improve throughput by processing
multiple PRs in parallel (optimistic vertical parallelization) or by
allowing the setup of multiple pipelines (static horizontal
parallelization), their overall throughput is constrained by window
sizes and PR success rates. These limitations can cause suboptimal
performance in fast-paced environments with fluctuating workloads or
high failure rates.

More tailored systems, as Uber's SubmitQueue~\cite{ananthanarayanan2019keeping},
implement a comprehensive approach to merge pipeline optimization using both
pessimistic vertical and dynamic horizontal parallelization based on predictive
modeling and conflict analysis on targets in the build graph. SubmitQueue
performs conflict analysis based on build targets to determine which PRs or
groups of PRs can be tested independently. Additionally, it uses machine
learning to speculate on the pruned paths within the merge pipeline that are
most likely to succeed. This probabilistic approach increases
throughput and minimizes resource waste by scheduling the paths that are likely
to succeed.  Uber further improved their system by allowing smaller PRs to
bypass larger ones, utilizing the commutativity of
builds~\cite{lin2023bypassing} and enhancing scheduling based on predicted build
times~\cite{juloori2025ci}.
AirBnB's Evergreen system also uses optimistic vertical parallelization in
combination with horizontal dynamic parallelization to test independent
changes concurrently~\cite{kudelka2022evergreen}.
Aviator, building on top of Uber's
SubmitQueue~\cite{ananthanarayanan2019keeping}, introduced a threshold to cut
off speculations that are not worth building~\cite{jain2023aviator}.

These sophisticated systems apply advanced techniques like predictive modeling
based on build tool features and conflict analysis. However, these approaches
rely heavily on strong assumptions about the build system, such as consistent
build graph structures and the ability to precisely analyze conflicts between
PRs. The assumption that such information is available may not hold
for more general or heterogeneous build systems, limiting the applicability of
such solutions. Furthermore, adopting Uber's approach may require a complete
overhaul of existing CI infrastructure, as it depends on a deep integration of a
sophisticated build systems. This reduces its accessibility for organizations that rely on standard CI tools and those unable to invest in significant infrastructure changes.

In contrast, the approach presented in this study achieves significant
improvements in throughput without requiring substantial changes to existing CI
systems. By leveraging predictive modeling using widely available,
language-agnostic, information, our method avoids relying on rigid assumptions
about the build system, making it more broadly applicable and easier to
integrate into standard CI practices.


\section{PRioritize: Predictive Prioritizations in Merge Pipelines}

To address the limitations (\eg deep integration of sophisticated build systems)
of existing merge pipeline optimizations, we propose a novel, practical approach
that reorders PRs within the merge pipeline in order to increase throughput by
prioritizing PRs that are likely to pass. This strategy, called \prioritize,
aims to optimize pipeline throughput during peak activity periods---the time
when high throughputs in merge pipelines are most needed---by dynamically
adjusting the scheduling of PRs. Rather than relying on static configurations or
requiring complete infrastructure overhauls, our approach integrates easily
into existing CI systems and focuses on improving efficiency during high
developer activity.

The core idea of our approach is to prioritize PRs with a higher probability of
success and deprioritize those likely to fail. By doing so, we reduce the risk
of pipeline resets and cascading delays, ensuring that successful changes are
integrated promptly. During peak developer activity, when the volume of incoming
PRs is highest, our system reserves pipeline capacity for PRs with a lower
failure likelihood. Waiting PRs are ranked based on their failure likelihood.
When $k$ spots become available in the merge pipeline, the $k$ waiting PRs with
the lowest failure likelihood are inserted into the merge pipeline. Typically,
merge pipelines complete one at a time, and the system checks for available free
slots. However, in scenarios where the window size increases or due to
infrastructure-related issues, it is possible for multiple builds to finish
simultaneously, resulting in more than one slot becoming available. By
prioritizing likely passing changes, the system reduces delays, improves
developer productivity, and minimizes disruptions caused by pipeline resets.
Conversely, PRs with higher failure likelihood scores are processed during
off-peak hours.

Likely failing PRs can of course be handled in different ways; in this
paper we reorder likely failing PRs to process them during off-peak
hours. This approach is straightforward, enables systematic
experimentation through simulations, and aligns with scenarios like
that of BMW, where peak and off-peak hours are
well-defined.
%
To avoid that developers of likely failing builds might experience
longer wait times, we propose investigating additional mechanisms,
such as giving developers the option to re-review their PRs in the
context of the updated mainline and resubmitting them to the pre-merge
CI pipeline, or using explainable AI methods to provide insights into
what causes PRs to be predicted to fail.  This would addressing
potential problems proactively, reducing delays in subsequent
processing cycles.  Importantly, even with poor model performance, the
worst-case scenario is not a broken main branch but merely longer
waiting times for some developers, since all changes are still
individually tested. This can be continuously monitored and evaluated,
allowing for quick response by disabling the approach and reverting.
By introducing \prioritize, we aim to establish a baseline for future research
in this field, such as on other means to treat likely failing PRs.

\begin{figure}
  \begin{center}
  \includegraphics[width=0.45\textwidth]{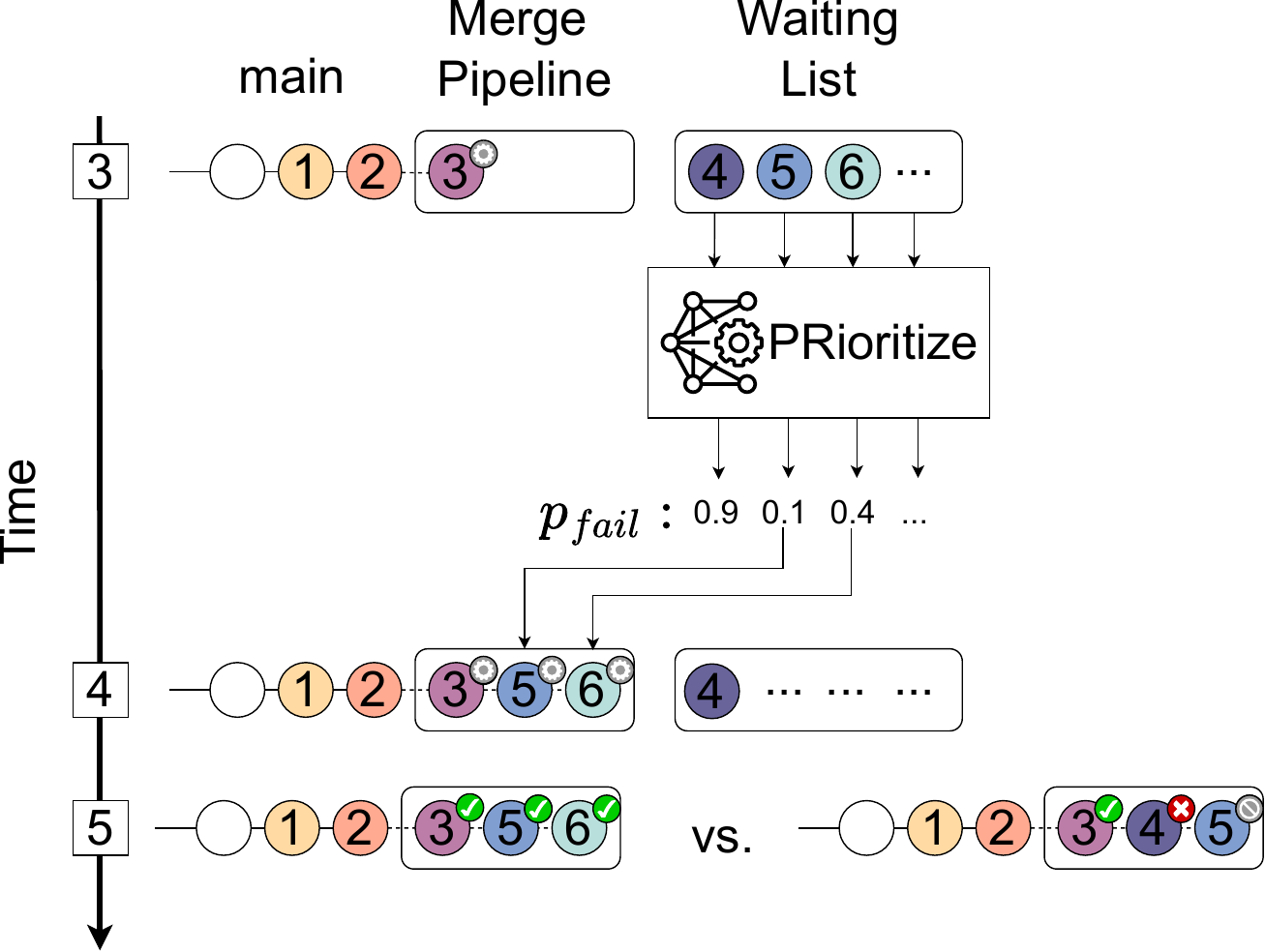}
  \end{center}
  \caption{Illustration of the \prioritize approach, showcasing its ability to
  optimize pipeline throughput by strategically selecting PRs based on failure
  likelihood scores.}
  \label{fig:prioritize}
\end{figure}

The advantages of the prioritization are exemplified in
\cref{fig:optimistic} at \insidebox{3} (time step three). At this
point, \textit{PR1} and \textit{PR2} successfully merge, creating two
available slots in the merge pipeline. As illustrated in
\cref{fig:prioritize}, \prioritize intercepts this process and
calculates failure likelihood scores for all pending PRs. Using these
scores, \prioritize inserts PRs that have a lower probability of
failure. In this scenario, it selects \textit{PR5} and \textit{PR6},
which are inserted at \insidebox{4} (as shown in
\cref{fig:prioritize}), but deprioritizes \textit{PR4}. By doing so,
\prioritize avoids unnecessary resets at \insidebox{5}. Consequently,
\textit{PR3}, \textit{PR5}, and \textit{PR6} successfully complete all
test steps and are ready for merging. This reduces cascading delays,
which would occur if the pipeline inserted PRs in default order
(\textit{PR4} and \textit{PR5}). A sequential insertion would lead to
a reset at \insidebox{5} (right), requiring the rescheduled processing
of \textit{PR5} and insertion of \textit{PR6} later. By optimizing the
sequence of PRs, \prioritize enhances the overall pipeline
throughput. Activating this method during business hours maximizes its
effectiveness, when there is a high volume of queued PRs to select
from.

While \textit{merge pipeline resets} may still occur during off-peak
hours, this timing minimizes their impact on productivity; developers
typically do not require fast feedback overnight, as their primary
concern is that their PRs are processed by the next business day. By
shifting resets to non-business hours, the system keeps peak hours
free for processing PRs that are more likely to succeed, ensuring
smoother and faster development workflows when developers are most
active. Our approach is designed to integrate incrementally with
existing CI systems. By augmenting current pipelines with predictive
modeling, organizations can achieve significant throughput
improvements without requiring extensive infrastructure changes. This
makes our method a practical and cost-effective solution.

To achieve this prioritization and calculate failure likelihood scores, we
leverage machine learning. While the pre-merge pipeline ensures that all PRs in
the merge pipeline individually build and pass their test stages, the merge
pipeline offers a different context, as PRs are not evaluated in isolation but
against the current state of the main branch. The predictions are therefore made
at the PR level, where the model estimates the likelihood of a PR causing a
failure when enqueued in the context of merge pipelines based on features
extracted from PR metadata and CI history of the individual PR. To enable the
model to learn this specific context and identify patterns of which PRs fail in
the merge pipeline, it requires training exclusively on historical data from the
merge pipeline itself.

\section{Implementation of PRioritize}
This section describes the implementation details of our predictive model for
prioritizing PRs in merge pipelines.

\subsection{Input Features}

Given the size, complexity, and diversity of codebases using merge
pipelines~\cite{lucido2017uber,jungwirth2025shifting,ananthanarayanan2019keeping},
it is important to use features that are both language-agnostic and easily
retrievable in large-scale systems. Our feature set can be broadly categorized
into two groups: PR-Metadata and CI-History.

\subsubsection{PR-Metadata}
This category includes features related to the PR itself, capturing the
characteristics of the changes being proposed. The features related to diffs,
such as changed lines or files, are relative to the current head of the main
branch when the PR was enqueued for processing in the merge pipeline.
\begin{itemize}
  \item \textit{\#additions}: Number of lines added in the PR.
  \item \textit{\#changed\_files}: Number of files changed in the PR.
  \item \textit{\#comments}: Number of comments in the PR.
  \item \textit{\#commits}: Number of commits in the PR.
  \item \textit{\#deletions}: Number of lines deleted in the PR.
  \item \textit{\#depends\_on}: Number of other PRs this PR depends on. In our
  setup, dependencies are explicitly specified using Zuul CI's dependency
  management feature for cross-repository changes in our virtual monorepo. This
  feature may not be available in traditional monorepos that handle such
  dependencies via single PRs.
  \item \textit{\#reviews}: Number of review comments in the PR; more discussion may be evidence of quality issues or controversial changes.
  \item \textit{changed\_files}: Cluster identifier of changed files in the PR.
  We used a Bag-of-Words model~\cite{harris1954distributional} to transform the
  file names into a vector. To reduce dimensionality and make the data suitable
  for machine-learning, we used a K-Nearest-Neighbor
  Classifier~\cite{fix1985discriminatory}.
\item \textit{is\_cyclic\_dependent}: Boolean indicating whether the
  PR is cyclic-dependent with another one due to BMW's virtual
  monorepo~\cite{jungwirth2025shifting}. Cyclic dependencies can
  impact systems, as they often indicate changes across project
  boundaries.
  \item \textit{pr\_created}: Time in hours since the PR was created; the longer a PR has been waiting, the larger the potential drift compared to the main branch.
\end{itemize}

\subsubsection{CI-History}
This category captures features related to the CI history of the PR,
allowing the model to learn from prior results. Specifically, we use
pre-merge CI history features, as PRs typically pass through the merge
pipeline only once (only if they fail they are enqueued a second
time). Consequently, features related to the merge pipeline would
often contain sparse or empty values due to the lack of historical
data. However, we also incorporate merge pipeline features, as they
help identify PRs that are likely to cause failures in the merge
pipeline. This combination of pre-merge CI history and merge pipeline
features proved to deliver the best performance in our approach.
\begin{itemize}
  \item \textit{\#files\_in\_prev\_resets}: Number of files modified in PRs that
  triggered a merge-queue failure.
  \item \textit{\#pre-merge\_failures}: Number of failed pre-merge pipeline runs
  for the PR.
  \item \textit{\#pre-merge\_runs}: Number of all pre-merge pipeline runs (passed
  and failed) for the PR.
  \item \textit{\#previous\_merge\_failures}: Number of previous
  merge pipeline failures for the PR.
  \item \textit{avg\_pre-merge\_duration}: Average duration of pre-merge pipeline
  runs for the PR.
  \item \textit{change\_queue}: Numerical encoding indicating in which merge
  pipeline the PR is enqueued. In CI systems employing multiple merge pipelines,
  failure patterns can vary across queues. This feature enables the model to
  capture such differences. Although training separate models for each pipeline
  is possible, we achieved a better performance using this discriminative
  feature, which allows the model discover general, as well as pipeline-specific
  patterns.
  \item \textit{last\_pre-merge\_failure}: Time since the last failed pre-merge
  pipeline run for the PR.
  \item \textit{last\_pre-merge\_run}: Time since the last pre-merge pipeline run
  for the PR.

\end{itemize}

We selected the features based on insights from our large-scale CI
systems, and insights from prior
research~\cite{machalica2019predictive,chen2020buildfast,sun2024ravenbuild}. For
example, features such as \textit{\#pre-merge\_failures} and
\textit{last\_pre-merge\_failure} draw inspiration from prior work
that emphasizes the importance of historical failure data in
predictive
modeling~\cite{machalica2019predictive,chen2020buildfast,sun2024ravenbuild}.
Similarly, PR-specific metadata such as \textit{\#changed\_files} and
\textit{\#additions} align with established practices in evaluating
the complexity and risk of
changes~\cite{machalica2019predictive,chen2020buildfast,sun2024ravenbuild}.
While the specific feature set reflects the implementation for our
case study, the features are language-agnostic and should be
applicable to other CI environments using merge pipelines.

\subsection{Training}
\label{sec:train}


Although some features relate to pre-merge pipeline data and changes
relative to an earlier state of the main branch, the training data
itself requires a representative dataset of historical CI merge
pipeline events, including labeled PR outcomes (we opt to label
failing PRs with 1 and failing with 0). These labels allow the model
to predict failure likelihood scores, which we then use to rank
waiting PRs. Since passing PRs outnumber failing ones, the dataset is
imbalanced. To address this class imbalance, one approach is to
include only passing PRs that were processed concurrently before a
failing PR in the same merge pipeline. Such passing PRs provide
contextual information, enabling the model to learn patterns that
distinguish between passing and failing PRs more effectively.

\subsection{Model}

To address the challenges of imbalanced datasets and the complexity of
scaling in continuous systems, it is important to use a model that is
robust against imbalanced data and minimizes preprocessing
requirements. Based on previous
research~\cite{chen2020buildfast,sun2024ravenbuild}, we therefore use
Gradient Boosted Decision Trees~\cite{chen2016xgboost}, implemented
via the Python library XGBoost~\cite{2022xgboost}, as it satisfies
these requirements.  XGBoost simplifies the preprocessing pipeline by
not requiring feature scaling~\cite{machalica2019predictive} and is
particularly effective for imbalanced data~\cite{chen2016xgboost}.


We tuned the hyperparameters of XGBoost using Bayesian optimization
with cross-validation (implemented in the bayesian-optimization Python
package~\cite{nogueira2014bayesian}), aiming to maximize the
Precision-Recall Area Under the Curve. The hyperparameters tuned
include: \textsf{max\_depth}, \textsf{min\_child\_weight},
\textsf{gamma}, \textsf{subsample}, \textsf{colsample\_bytree},
\textsf{learning\_rate}, \textsf{reg\_alpha}, and
\textsf{reg\_lambda}.
The general methodology can be applied to other contexts as
well.



\section{Case Study}
To evaluate our approach, we conducted a case study within BMW using CI replays to simulate real-world scenarios without impacting the production environment. This aim of this study is to address the following research questions:
\begin{description}
  \item[RQ1 (Snapshots):] How well can \prioritize delay failing PRs?
  \item[RQ2 (Simulation):] How well can \prioritize increase throughput in merge
  pipelines during business hours?
\end{description}

\subsection{Monorepo and CI Infrastructure at BMW}
\label{sec:ci_infra}


BMW employs a virtual monorepo at the driving dynamics and autonomous driving
departments that consolidates several submodules and third-party dependencies
into a single workspace. This monorepo serves as the foundation for our driving
dynamics and autonomous driving software. The virtual monorepo approach is
adopted due to legal restrictions and to limit visibility across our
collaboration partners. It comprises approximately 80 million lines of code
spanning multiple programming languages, with C++ being the predominant choice
for in-car software and Python for
tooling~\cite{jungwirth2025shifting,schwendner2025practical}.

We use Zuul~\cite{Zuul} as our CI system. Zuul employs the concept of
pipelines, which represent workflows consisting of jobs planned for
execution. A pipeline is triggered by specific events, with each job
performing a distinct task such as running a test suite. Our pipelines
are executed at various points in the development lifecycle. Pre-merge
pipelines run fast, stable tests tailored for each project scope to
provide quick feedback before changes are merged. Merge pipelines
ensure compatibility between concurrent changes, while post-merge
pipelines execute time-consuming or flaky tests after changes are
integrated into the main branch. Periodic pipelines perform scheduled
tasks, such as dependency management and monitoring.

\subsection{Merge Pipelines at BMW}

To handle the scale of development activities, we use Zuul's vertical
optimistic parallelization with a contracting window of size
7--25. When a failure occurs we halve the window size and for every
successful merge we increase it by one. We also apply horizontal
scaling through three independent merge pipelines across our driving
dynamics and autonomous driving software stack, each pipeline serving
a specific project scope. Overnight, these pipelines are consolidated
into a single pipeline to merge dependent PRs across all
pipelines. The separation of merge pipelines was introduced to
increase throughput while addressing scaling limits and maintaining
confidence in identifying breakages across the stack.

\begin{figure}
  \centering
  \includegraphics[width=0.45\textwidth]{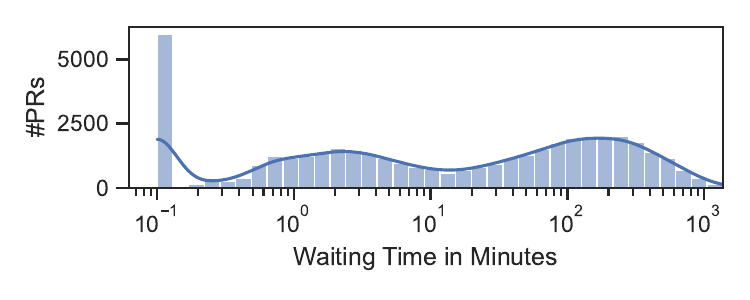}
  \vspace{-1em} 
  \caption{Distribution of waiting times of PRs in the merge pipelines at BMW.}
  \label{fig:waiting}
\end{figure}

On average, over 400 changes are merged daily across our merge pipelines,
highlighting the scale and complexity of our software integration process.
Currently, PRs in our pipelines experience an average of 50 minutes of waiting
time before processing begins (cf.~\cref{fig:waiting}), in addition to
60 minutes of processing time. This indicates a significant opportunity for
optimization, particularly during business hours when developers are actively
waiting for their changes to be merged. Improved scheduling during these hours
could reduce delays, enabling developers to iterate faster and increase
productivity.

BMW has defined business hours (noticeable in~\cref{fig:pr_over_day})
because the geographical spread of its developers is limited, which
allows for targeted optimizations during peak activity
periods. Outside of business hours, the waiting time is less critical,
as the primary concern for developers is whether their PRs are
successfully merged by the time they return. Addressing inefficiencies
during peak hours remains a priority to ensure seamless integration
and minimize disruptions to the workflow. This underscores the need
for smarter scheduling to enhance the efficiency of our merge
pipelines and better accommodate the demands of high-velocity
development environments.

\subsection{Dataset}
To create our training dataset, we logged merge pipeline activities
over a span of approximately four months (2024-10-22 to 2025-02-23)
covering 89 days. We excluded 16 days with major infrastructure
issues, as well as the period between the Christmas holidays
(2024-12-20 to 2025-01-08), as there was no active development during
this time. The resulting dataset contains 670 failing and 4000 passing
builds from our three merge pipelines (see~\cref{tab:dataset}). We
included only the passing builds that were executed before and in
parallel with a failing one in a single merge pipeline, as outlined
in~\cref{sec:train}. Additionally, we treated rescheduled PRs after a
failure as not failed (because they did not cause the failure);
instead, we evaluated their outcome based on the next
iteration. Failures due to dependencies on a failing PR were filtered
out.

The test dataset spans 15 business days, from 2025-02-24 to 2025-03-14. The
split between training (89 days) and test data (15 business days) was chosen
based on temporal considerations rather than a standard percentage split. The
89-day training period ensures sufficient samples for machine learning,
particularly given the class imbalance of fewer failing than passing PRs. The
15-day test period represents a recent, representative evaluation window that
maintains realistic experimental conditions while providing fresh data for
validation. This temporal approach better reflects real-world deployment
scenarios where models train on historical data and predict future outcomes. To
ensure robustness and prevent data leakage. This approach respects the temporal
order of events, ensuring that the test dataset reflects the natural evolution
of the system and does not include information from the future during model
training. No further filtering was used in order to simulate real-world
deployment scenarios. It contains 506 failing PRs and 7221 passing PRs in the
merge pipeline.

To address RQ1, we reconstructed snapshots of the merge pipeline, including the
currently processed PRs and the waiting list, which allows us to compare our
approach against the baseline in a pointwise manner. For RQ2, we utilized all
PRs to simulate the merge pipeline and demonstrate whether our approach
increases the throughput during business hours.
\begin{table}
  \caption{Characteristics of the train and test dataset used in the case study.
  (*The testset contains only data from business days within the given time
  period.)}
  \begin{center}
  \begin{tabular}{lrrrrr}
  \toprule
  & \textbf{FAILED} & \textbf{PASSED} & \textbf{START} & \textbf{END} & \textbf{DAYS} \\
  \midrule
  \textbf{Train} & 670 & 4000 & 2024-10-22 & 2025-02-23 & 89 \\
  \textbf{Test*} & 506 & 7221 & 2025-02-24 & 2025-03-15 & 15 \\
  \bottomrule
  \end{tabular}
  \end{center}
  \label{tab:dataset}
\end{table}

\subsection{Baselines}
In this study, we compare the performance of our proposed predictive
PR ordering method against several non-learning-based reordering
baselines, designed using our domain knowledge and practical
assumptions about factors that influence the likelihood of our PRs
pass in our merge pipelines.

The first baseline orders the waiting PRs in ascending order along the
number of pre-merge failures (\textit{\#pre-merge\_failures}),
prioritizing PRs submitted by developers with a lower historical
frequency of pre-merge failures.  The intuition behind this approach
is that PRs associated with fewer failures are more likely to succeed
and, therefore, should be processed earlier to minimize the risk of
pipeline resets. The second baseline prioritizes PRs based on the
number of changed lines (\textit{\#changed\_lines}) with the rationale
that smaller PRs are less complex and less likely to introduce
errors. Similarly, the third baseline uses the number of changed files
(\textit{\#changed\_files}) as a criterion, assuming that PRs
affecting fewer files are less prone to conflicts or failures. The
last baseline considers the time since the last pre-merge failure
(\textit{last\_pre-merge\_failure}), with the idea that PRs submitted
by developers with a longer gap since their last failure are more
likely to succeed.

These baseline approaches are evaluated alongside our proposed method
(\prioritize), which employs a predictive model. The model leverages historical
build data, PR metadata, and contextual information to estimate the likelihood
of a successful build for each PR and dynamically optimizes their ordering in
the merge pipeline. While we expect the predictive model to outperform the
baseline heuristics due to its ability to combine and utilize multiple factors
simultaneously, we include these baselines to demonstrate the value of the
additional effort required to implement \prioritize.

\subsection{Evaluation and Simulation Environment}
To answer RQ1, we evaluate the effectiveness of the \prioritize
approach in deprioritizing failing PRs using 506 snapshots of the
merge pipeline each containing one of the failing PRs from our test
dataset. Each snapshot represents the state of the pipeline at the
moment a failure occurs.  These snapshots include both the PRs
currently being processed and those waiting in the queue. Since merge
pipelines operate as continuous systems where PRs are enqueued
whenever developers consider them ready for merging, the arrivals of
PRs follow a largely chaotic pattern. With thousands of developers
working independently without coordinating every submission decision,
the order in which PRs enter the merge pipeline is mostly left to
chance rather than being consciously controlled. While there can be
dependencies across PRs submitted by developers, such as shared code
changes or sequential updates, which influence the order and timing of
their arrival, the overall submission pattern remains largely
uncoordinated. To account for this, we simulate a FIFO baseline at
each snapshot by randomly shuffling the PRs \num{10000} times. This
simulated FIFO baseline provides a reference for comparing \prioritize
and other heuristic-based approaches. We use the simulated \num{10000}
FIFO datapoints for each snapshot to calculate the \textit{absolute
  failure rank}, which is the average absolute rank of the failing PR
across the \num{10000} permutations for each snapshot. For the
baseline heuristics, the absolute failure rank is determined by
sorting the PRs based on the given method in each snapshot and
reporting the absolute rank of the failing PR. To account for the
random nature of learning based approaches, we assessed \prioritize
absolute failure ranks for each snapshot averaged over ten times
including retraining. To measure the effectiveness of each approach,
we calculate the delay by subtracting the absolute failure rank of
\prioritize and the heuristic baselines from the simulated FIFO
absolute failure rank for each snapshot. This comparison allows us to
evaluate how effectively the different approaches delay failing
PRs. To assess statistical significance, we apply the Wilcoxon
signed-rank test~\cite{woolson2005wilcoxon} at $\alpha = 0.05$ to
determine whether the absolute ranks achieved by \prioritize and the
heuristic baselines are significantly larger than those of FIFO across
the 506 snapshots.  Additionally, we use the Vargha-Delaney
A$_{12}$~\cite{vargha2000critique} effect size.

For RQ2, we simulate the real-world performance and throughput impact
of our approach during our business hours (9:00 AM to 5:00 PM) to
assess its effectiveness in improving throughput during high-peak
periods. While the snapshot-based evaluation provides insights into
prioritization performance, it does not fully capture the practical
benefits in dynamic, real-world scenarios.  The simulation replicates
the behavior of our three merge pipelines, incorporating critical
factors such as PR outcomes, interdependencies, scheduling
constraints, and execution times. This setup models the processing of
PRs under varying workloads, specifically evaluating scenarios with
window sizes of 5, 10, 15, and 20 waiting PRs.  For each of the 15
days covered by the test dataset, we enqueue PRs in the order they
actually arrived on the day. Whenever a PR is completed by the
simulated merge pipelines, the next PR to be processed is selected
from the window of succeeding PRs, ranked using the actual order,
\prioritize, or one of the baselines. If a failing PR is processed,
then the simulation replicates the actual reset behavior of the merge
pipelines.

Since the simulation only considers business hours (8 hours) but the
test dataset contains the PRs of the entire day, we can observe
variations in throughput per day, quantified as the number of merged
PRs across our three merge pipelines for each of the 15 days.  To
account for the random nature of \prioritize we averaged over ten
retries of the simulation including retraining. The statistical
significance of throughput improvements is assessed using the Wilcoxon
signed-rank test~\cite{woolson2005wilcoxon} at $\alpha = 0.05$.
Specifically, we apply the test to evaluate whether the daily
throughput of the 15 simulated days, measured as the number of merged
PRs across the three merge pipelines, is significantly higher for
\prioritize and the heuristics compared to the baseline FIFO
approach. Additionally, we use the Vargha-Delaney
A$_{12}$~\cite{vargha2000critique} effect size.

\subsection{Threats to Validity}
\subsubsection{External Validity} The
generalizability of our findings is limited by the specific industrial context
of BMW. While our CI and code infrastructure may be representative for other
large-scale automotive software-systems, the results might not generalize to
other domains. We gave detailed insights into our context and refrained from
using context-specific features, relying on readily available features in
large-scale systems.

\subsubsection{Construct Validity} The pointwise comparison on merge pipeline
snapshots (RQ1) quantifies effectiveness under controlled conditions, but may
not reflect real-world improvements. We also conducted a realistic simulation
(RQ2) replicating merge pipeline behavior, including PR processing times and
dependencies. Combining both analyses helps mitigate this threat to validity.

\subsubsection{Internal Validity}
Our approach relies on historical build data, PR metadata, and execution logs,
which may contain inaccuracies due to system errors or manual interventions. We
validated the data by excluding days with major infrastructure issues and
monitored for significant changes during the study period to reduce bias.


\section{Results}
This section outlines the results using snapshots (RQ1) and the
simulation of our merge pipelines (RQ2).

\subsection{RQ1: Delaying Likely Failing PRs in Snapshots}
\begin{figure}
  \centering
  \includegraphics[width=.45\textwidth]{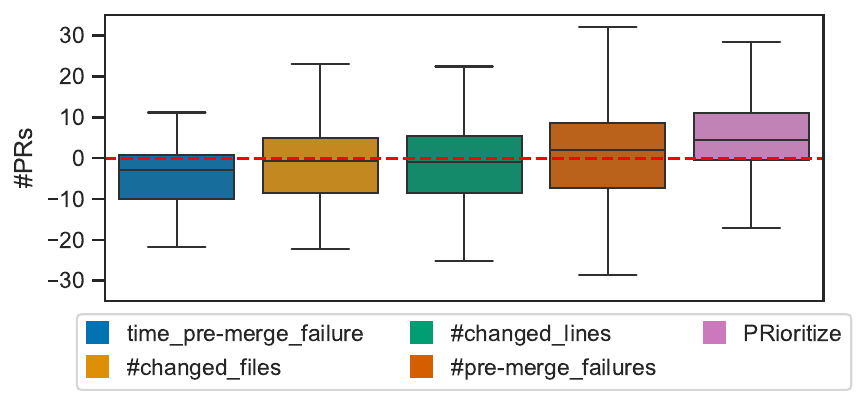}
  \caption{Distribution of the delays of failing PRs in number of PRs per merge
  pipeline snapshot compared against the simulated FIFO baseline (red dashed
  line).}
  \label{fig:re_perf}
\end{figure}

To assess the effectiveness of our model in delaying failing PRs, we evaluate
its performance using snapshots of our merge pipeline, comparing the delay in
PRs per merge-pipeline failure against a simulated FIFO baseline.

\Cref{fig:re_perf} shows that \prioritize achieves the highest delays,
with a mean of 5.94 PRs and a median of 4.45 PRs, outperforming all
other approaches. \prioritize exhibits the smallest interquartile
range of 11.64 PRs, indicating more consistent performance across
different scenarios.  Importantly, \prioritize is the only approach
that significantly outperforms the FIFO baseline, as evidenced by an
effect size of $A_{12} = 0.69$ and $p <0.01$. In contrast, other
heuristics either fail to delay failing PRs or perform worse than the
FIFO baseline. For example, \textit{time\_pre-merge\_failure} has a
mean delay of -5.40 PRs, a median delay of -2.96 PRs performing worse
than FIFO. The effect size ($A_{12} = 0.36$) indicates a small effect
size, with $p = \sim1.0$, confirming no significant improvement over
FIFO. Similarly, \textit{\#changed\_files} and
\textit{\#changed\_lines} perform poorly, with mean delays of -3.25
PRs and -1.76 PRs, respectively, and median delays of -0.76 PRs and
-0.90 PRs with negligible effect sizes (0.46 and 0.52) and no
statistical significance ($p>0.05$).  \textit{\#pre-merge\_failures}
exhibits a slight improvement over the FIFO baseline, with a mean
delay of 1.88 PRs and a median delay of 1.90 PRs, alongside an effect
size of 0.53. However, the improvement is negligible and not
statistically significant ($p>0.05$). These results highlight that
\prioritize significantly outperforms all other approaches, achieving
the highest and most consistent delays for failing PRs while
demonstrating meaningful improvements over the FIFO baseline.

\summary{1}{\prioritize significantly outperforms FIFO, achieving the
  highest average delay of failing PRs (5.94 PRs) with a medium effect
  size ($A_{12} = 0.69$). Other heuristics perform worse or show
  negligible improvements over FIFO.}


\subsection{RQ2: Throughput Improvements in the Simulation}

\begin{table}[t]
  \begin{center}
    \caption{Average number of merged PRs per day across our three merge-queues. Statistically significant differences to the FIFO baseline ($p < 0.05$) are marked in bold.}
    \begin{tabular}{lrrr}
      \toprule
      & \textbf{\#merged} & \textbf{compared} & \textbf{Vargha-} \\
      & \textbf{PRs} & \textbf{to FIFO} & \textbf{Delaney} \\
      \midrule
\textbf{Waiting PRs: 5} & \\
 \midrule
      \textbf{\prioritize} & \textbf{206} & \textbf{+7} & \textbf{0.59} \\
      \textit{\#pre-merge\_failures} & 205 & +6 & 0.56 \\
      \textit{\#changed\_lines} & 187 & -12 & 0.37 \\
      \textit{\#changed\_files} & 191 & -8 & 0.35 \\
      \textit{last\_pre-merge\_failure} & 189 & -10 & 0.38 \\
      \midrule
\textbf{Waiting PRs: 10} & \\
 \midrule
      \textbf{\prioritize} & \textbf{212} & \textbf{+13} & \textbf{0.59} \\
      \textit{\textbf{\#pre-merge\_failures}} & \textbf{210} & \textbf{+11} & \textbf{0.60} \\
      \textit{\#changed\_lines} & 189 & -10 & 0.37 \\
      \textit{\#changed\_files} & 189 & -10 & 0.32 \\
      \textit{last\_pre-merge\_failure} & 187 & -12 & 0.35 \\
      \midrule
\textbf{Waiting PRs: 15} & \\
 \midrule
      \prioritize & 211 & +12 & 0.63 \\
      \textit{\#pre-merge\_failures} & 206 & +7 & 0.58 \\
      \textit{\#changed\_lines} & 183 & -16 & 0.29 \\
      \textit{\#changed\_files} & 190 & -9 & 0.38 \\
      \textit{last\_pre-merge\_failure} & 182 & -17 & 0.27 \\
      \midrule
\textbf{Waiting PRs: 20} & \\
 \midrule
      \textbf{\prioritize} & \textbf{212} & \textbf{+13} & \textbf{0.66} \\
      \textit{\#pre-merge\_failures} & 207 & +8 & 0.57 \\
      \textit{\#changed\_lines} & 179 & -20 & 0.24 \\
      \textit{\#changed\_files} & 189 & -10 & 0.36 \\
      \textit{last\_pre-merge\_failure} & 182 & -17 & 0.28 \\
      \bottomrule
    \end{tabular}
    \label{table:mergedPRsComparison}
  \end{center}
\end{table}

\begin{figure*}
  \centering
  \includegraphics[width=\textwidth, trim=0 0.6cm 0 0]{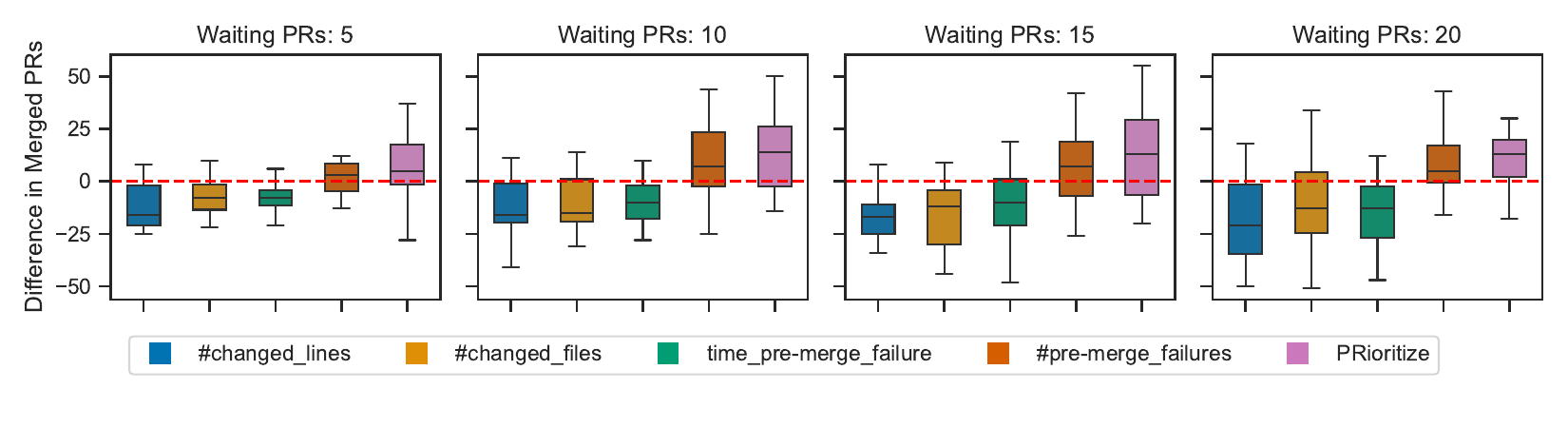}
  \vspace{-2em} 
  \caption{Difference in average number of merged PRs per day compared to the FIFO baseline (red
  dashed line) for each simulated window of waiting PRs (5, 10, 15, and 20).}
  \label{fig:diff_box}
\end{figure*}

To evaluate the real-world performance of \prioritize (\ie the
increase in throughput that it creates), we use a simulation of our
merge pipelines to assess its impact on the number of merged PRs
compared to heuristic-based baselines.

\Cref{table:mergedPRsComparison} shows the total number of merged PRs
for the \prioritize model and the baseline heuristics. Across all the
waiting PR windows (5, 10, 15, and 20), \prioritize consistently
achieves the highest number of merged PRs, merging up to 212 PRs on
average per day, compared to the FIFO baseline's 199 PRs. The
\textit{\#pre-merge\_failures} heuristic performs second-best, merging
up to 207 PRs, but other heuristics such as \textit{\#changed\_lines}
and \textit{time\_pre-merge\_failure} perform poorly, merging fewer
PRs than the baseline. The boxplot in \cref{fig:diff_box} shows the
difference in the number of merged PRs compared to the
baseline. \prioritize demonstrates statistically significant
improvements, with $p$-values ranging from 0.02 to 0.06, and effect
sizes up to $A_{12} = 0.66$. Heuristic-based baselines offer only
slight improvements, with effect sizes ranging from 0.56 for
\textit{\#pre-merge\_failures} to 0.24 for \textit{\#changed\_lines}.

In addition to its overall performance, \prioritize demonstrates effective
scaling across varying workloads. As the number of waiting PRs increases, the
difference in performance between \prioritize and the FIFO baseline becomes
more pronounced. For instance, with 20 waiting PRs, \prioritize merges 212
PRs, which is 13 more than FIFO, while with only 5 waiting PRs, the improvement
is 7 PRs. This trend indicates that \prioritize benefits from more
waiting PRs, as it has more PRs to prioritize and optimize. In contrast,
heuristics such as \textit{\#changed\_lines} and
\textit{time\_pre-merge\_failure} often perform worse as the number of waiting
PRs increased.

These results demonstrate that \prioritize scales effectively
across varying workloads and maximizes throughput during business hours, where
developer productivity is critical.

\summary{2}{The results of the simulation show that \prioritize
significantly improves throughput by up to 13 PRs in merge pipelines during
business hours per day. This represents a \perc{7} improvement compared to the FIFO
baseline.}


\section{Related Work}
\subsection{Merge Pipeline Systems}
General-purpose merge pipeline systems, such as GitHub's merge
queue~\cite{gh-merge}, GitLab's merge train~\cite{gl-merge}, and Zuul gate
queue~\cite{Zuul}, support a wide range of software with optimistic vertical and
static horizontal parallelization. However, they might face throughput
challenges in high-demand environments with high failure rates. These
limitations motivated us to enhance these systems, improving throughput without
significant infrastructure changes.
More advanced systems, like Uber's
SubmitQueue~\cite{ananthanarayanan2019keeping}, address these challenges using
pessimistic vertical and dynamic horizontal
parallelization with deep integration of
sophisticated build systems~\cite{ananthanarayanan2019keeping,
juloori2025ci,jain2023aviator,kudelka2022evergreen}. Uber further enhanced their system by exploiting
the commutativity of merging PRs~\cite{lin2023bypassing} and better scheduling
by runtime prediction~\cite{juloori2025ci}. While effective, such methods rely
on rigid assumptions and specialized infrastructure, limiting their
applicability. In contrast, our approach uses widely available,
language-agnostic information, enabling practical and scalable improvements
within standard CI environments.

\subsection{Build Result Prediction}
While defect prediction~\cite{fenton1999defect, ambros2012defect, tan2015defect,
madeyski2017defect, wan2020defect} primarily focuses on identifying defects in
source files, build failures in CI systems can arise from a variety of changes
beyond source code modifications (e.g., updating the version of a dependency).
As a result, defect prediction techniques cannot be directly applied to build
failure prediction~\cite{chen2020buildfast}.
Therefore, various techniques have been applied to predict build outcomes,
including data stream mining~\cite{finlay2014stream} and evolutionary search
algorithms~\cite{saidini2020evo}. However, machine learning techniques remain
the most commonly used approach~\cite{sun2024ravenbuild, chen2020buildfast,
jin2020cost, hassan2017changeaware, kamath2024combining, jin2017effective,
jin2023hybridcisave, abdalkareem2021skip}. These models leverage a diverse set
of features extracted from sources such as version control systems and
historical build outcomes~\cite{sun2024ravenbuild,chen2020buildfast}.

Building on prior work, we tailored our feature set and labeling
specifically to predict build outcomes in the context of merge
pipelines, as PRs are known to build in isolation since they already
passed through the pre-merge pipeline.


\section{Conclusions and Future Work}

Based on our analysis of existing merge pipeline systems and their
optimizations,
we proposed \prioritize, a practical approach that leverages build
outcome prediction to delay likely failing PRs outside of business
hours. 
Based on the encouraging results of our evaluation, we are now working
on enabling this approach in production, aiming to further validate
its impact and benefits to real-world development workflows.

The build prediction in our approach can also be used to further
optimize merge pipelines, such as by batching PRs predicted to pass,
allowing the pipeline to execute one build step for multiple changes.
Since merge pipelines typically create speculative integration states,
the build prediction might be further improved by incorporating
predictions for the exact states that will run in the pipeline.
To support further research and development in this area, we have made our
implementation publicly available at: \url{http://doi.org/10.6084/m9.figshare.29136350}


\bibliographystyle{IEEEtran}
\balance
\bibliography{IEEEabrv,bib/literature.bib}

All online resources accessed on 2025--05--13.

\end{document}